\documentclass{ws-procs975x65}            % default, citations in superscript

\def\beq{\begin{equation}}
\def\eeq{\end{equation}}
\def\bea{\begin{eqnarray}}
\def\eea{\end{eqnarray}}
\def\bealn{\begin{eqnarray}}
\def\eealn{\end{eqnarray}}

\def\ifb{\rm fb^{-1}}

\def\tev{\rm TeV}
\def\zp{{Z^\prime}}	% Z'
\def\dcol{D_{\rm{col}}}
\def\mres{M_{\rm{res}}}

\begin{document}
\title{Separating Dijet Resonances\\
Using the Color Discriminant Variable}

\author{E. H. Simmons$^\#$\footnote{Dr. Simmons presented this talk at SCGT15.}, R. S. Chivukula, P. Ittisamai, and N. Vignaroli}

\address{Department of Physics and Astronomy, Michigan State University,\\
East Lansing, MI, 48825, USA\\
$^\#$E-mail: esimmons@msu.edu}

\begin{abstract}
Color-singlet and color-octet vector bosons predicted in theories beyond the Standard Model have the potential to be discovered as dijet resonances at the LHC. A color-singlet resonance that has leptophobic couplings needs further investigation to be distinguished from a color-octet one. In previous work, we introduced a method for discriminating between the two kinds of resonances when their couplings are flavor-universal, using measurements of the dijet resonance mass, total decay width and production cross-section.  Here, we describe two extensions of that work.  First, we broaden the method to the case where the vector resonances have flavor non-universal couplings, by incorporating measurements of the heavy-flavor decays of the resonance.  Second, we apply the method to separating vector bosons from color-octet scalars and excited quarks.
\end{abstract}


\bodymatter

\section{Introduction}\label{sec:intro}

There have been many collider searches for Beyond-the-Standard-Model resonances decaying to dijet final states. The current LHC run will be able to detect such resonances out to larger masses. The question that we are pursuing in this talk is: once a new dijet resonance has been discovered, what can we deduce about it using information readily available after the discovery?

In previous work~\cite{Atre:2013mja}, we introduced a way to distinguish whether a vector resonance is a leptophobic $Z'$ or a coloron, using a construct that we called a ``color discriminant variable'', $\dcol$. The variable is constructed from the dijet cross-section for the resonance ($\sigma_{jj}$), its mass ($M$), and its total decay width ($\Gamma$), observables that will be available from the dijet channel measurements of the resonance:
	\beq
		\dcol \equiv \frac{M^3}{\Gamma} \sigma_{jj},
	\label{eq:dcol}
	\eeq
For a narrow-width resonance, the color discriminant variable is independent of the resonance's overall coupling strength.  We also illustrated applications of the color discriminant variable technique for two simple cases~\cite{Atre:2013mja}. The first was a flavor universal model with identical couplings to all quarks. In the second, the overall strength of couplings to quarks in the third generation was allowed to be different from those in the first two (couplings to top and bottom were kept equal).  While these two scenarios clearly illustrate the application of the method, they did not encompass the features of a typical $\zp$, whose up- and down-type couplings are usually also different from one another.

In this talk we extend the $D_{col}$ method to more general scenarios where couplings to quarks within the same generation (e.g., up vs. down, left-handed vs right-handed) are different~\cite{Chivukula:2014npa}.  We also show that the method may be used to distinguish among resonances decaying to $q\bar{q}$ (colored vector boson), $qg$ (excited quark) or $gg$ (color-octet scalar) final states~\cite{Chivukula:2014pma}.

%%%%%%%%%%%%%%%%%%%%%%%%%%%%
%***************************
%***************************
%%%%%%%%%%%%%%%%%%%%%%%%%%%%
\section{Vector Resonances: General Parameterization and Assumptions}
\label{sec:genpar}

 A coloron ($C$) or a $\zp$ that manifests as a dijet resonance is produced at hadron colliders via quark-antiquark annihilation. The interaction of a $C$ with the SM quarks $q_i$ is described by
	\beq
		\mathcal{L}_C  = i g_{QCD} C_\mu^a \sum_{i=u,d,c,s,t,b}
		\bar{q}_i\gamma^\mu t^a \left( g_{C_L}^i P_L + g_{C_R}^i P_R \right) q_i , \\
	\label{eq:colcoupl}
	\eeq
where $t^a$ is an $ \text{SU}(3) $ generator, $g_{C_L}^i$ and $g_{C_R}^i$ denote left and right chiral coupling strengths, relative to the strong coupling $g_{QCD}$, of the color-octet to the SM quarks. The projection operators have the form $P_{L,R} = (1 \mp \gamma_5)/2$ and the quark flavor index $i$ runs over $ i=u,d,c,s,t,b.$ Similarly, the interactions of a leptophobic $\zp$ with the SM quarks are given by
	\beq
		\mathcal{L}_{\zp}  = i g_w  \zp_\mu \sum_{i=u,d,c,s,t,b}
		\bar{q}_i \gamma^\mu\left( g_{\zp_L}^i P_L + g_{\zp_R}^i P_R \right) q_i,
	\label{eq:zpcoupl}
	\eeq
where $g_{\zp_L}^i$ and $g_{\zp_R}^i$ denote left and right chiral coupling strengths of the leptophobic $\zp$ to the SM quarks relative to the weak coupling $g_w = e/\sin\theta_W$.

Couplings between the vector boson and the left- and right-handed forms of the up- and down-type fermions need not be equal in general -- and the couplings can generally be different among the three generations of quarks. However, the observed suppressions of flavor-changing neutral currents disfavor a TeV-scale resonance with non-universal couplings to the first two generations~\cite{Bona:2007vi}; we therefore study only scenarios in which the couplings for the first two generations are identical.   With this in mind, a coloron would have six free parameters describing its couplings to quarks of different flavors, generations, and chiralities, while a leptophobic $Z'$ boson would have eight.  However, the observables of interest to us (width, mass, dijet cross-section) are insensitive to the chiral structure of the couplings. So we may denote
	\beq
		g^{q2} \equiv g_{L}^{q2} + g_{R}^{q2}
	\eeq
and conclude that there are only four relevant quark couplings for a coloron or $Z'$: 
	\bea
		g^{u\,2}_{C} &=& g^{c \,2}_{C},\quad g^{d\,2}_{C}=g^{s \,2}_{C},\quad g^{t\,2}_{C},\quad g^{b\,2}_{C}\\
		g^{u\,2}_{\zp} &=& g^{c \,2}_{\zp},\quad g^{d\,2}_{\zp}=g^{s \,2}_{\zp},\quad g^{t\,2}_{\zp},\quad g^{b\,2}_{\zp}~.
	\eea

Finally, in discussing a dijet cross section ($\sigma(pp \to V \to jj)$), we classify quarks from the first two generations as those yielding final-state jets:  $j = u,\, d,\, c,\, s$.  The cross sections involving decays to $t\bar{t}$ and $b\bar{b}$ final states (respectively, $\sigma_{tt} \equiv \sigma(pp \to V \to t\bar{t})$ and $\sigma_{bb} \equiv \sigma(pp \to V \to b\bar{b})$) will provide separate and valuable information. 

%%%%%%%%%%%%%%%%%%%%%%%%%%%%%%%%%%%%%%%%
%%%%%%%%%%%%%%%%%%%%%%%%%%%%%%%%%%%%%%%%
\section{The Color Discriminant Variable}
\label{sec:coldis}

%%%%%%%%%%%%%%%%%%%%%%%%%%%%
%***************************
%***************************
%%%%%%%%%%%%%%%%%%%%%%%%%%%%
\subsection{Review of the flavor universal scenario}
\label{subsec:flav-univ}
We first review the color discriminant variable in a flavor-universal scenario.  We work in the limit ($\Gamma/M \ll 1$) where the narrow-width approximation applies
	\beq
		\sigma_{jj}^V \equiv \sigma(pp \rightarrow V \rightarrow jj) \simeq \sigma(pp \to V) Br(V \to jj),
	\label{eq:genxsec}
	\eeq
where $\sigma(pp \to V)$ is the cross section for producing the resonance. Note that $Br(V \to jj)$ is the boson's dijet branching fraction, which equals $4/6$ for a flavor universal vector resonance that is heavy enough to decay to top quarks. In this limit, the total decay width for a heavy coloron or a $Z'$ is	
\beq
		\Gamma_C = \frac{\alpha_s}{2} M_C g_C^2, \quad\quad \Gamma_\zp = 3 \alpha_w M_\zp g_\zp^2\,,
	\label{eq:zpwiduniv}
	\eeq
where $g_{C/\zp}^2 = \left( g_{{C/\zp}_L}^2 + g_{{C/\zp}_R}^2\right)$ denotes the flavor-universal coupling of the resonance to quarks. These, respectively, lead to the dijet cross sections:
	\beq
		\sigma_{jj}^C = 	\frac{8}{9} \frac{\Gamma_C}{M_C^3} \sum_{q} W_q(M_C)  Br(C \to jj), \quad\quad 
		\sigma_{jj}^\zp 	= 	\frac{1}{9} \frac{\Gamma_{\zp}}{M_{\zp}^3} \sum_{q} W_q(M_\zp)  Br(\zp \to jj)\,.
	\label{eq:zpxsec}
	\eeq
Here the function $W_q$, which is constructed from the parton luminosity for the production of the vector resonance with mass $M_V$ via $q\bar{q}$ annihilation at the center-of-mass energy squared $s$, is defined by
	\beq
		W_{q}(M_V) = 2\pi^2 \frac{M_V^2}{s} \int_{M_V^2/s}^{1} \frac{dx}{x}
			\left[ f_q\left(x, \mu_F^2\right) f_{\bar{q}}\left( \frac{M_V^2}{sx}, \mu_F^2 \right) +
			f_{\bar{q}}\left(x, \mu_F^2\right) f_q\left( \frac{M_V^2}{sx}, \mu_F^2 \right) \right]  \,,
	\label{eq:w_function}
	\eeq
where $f_{q}\left(x,\mu_F^2\right)$ is the parton distribution function at the factorization scale $\mu_F^2$. Throughout this article, we set $\mu_F^2 = M_V^2$.

The color discriminant variables are then
	\bea
		\dcol^C &=&  \frac{M_C^3}{\Gamma_C} \sigma_{jj}^C= \frac{8}{9} \left[\sum_{q} W_q(M_C)  Br(C \to jj) \right]
	\label{eq:coldcol}\\
		\dcol^\zp &=& \frac{M_\zp^3}{\Gamma_\zp} \sigma_{jj}^\zp= \frac{1}{9} \left[\sum_{q} W_q(M_\zp)  Br(\zp \to jj)\right]
	\label{eq:zpdcol}
	\eea
for the coloron and $\zp$, respectively.  The factors in the square brackets in (\ref{eq:coldcol}) and (\ref{eq:zpdcol}) are the same for flavor-universal resonances having a particular mass; only the initial numerical factors differ. In other words, the different values of the color discriminant variables for the the two types of flavor-universal resonance
	\beq
		\dcol^C = 8 \dcol^\zp
	\label{eq:factor8_dcol}
	\eeq
will help pinpoint the nature of the color structure of the discovered particle.

\subsection{Flavor non-universal scenario}
\label{subsec:flavnon-univ-gen}
In a flavor non-universal scenario, we still follow the parameterization and assumptions introduced above.  The production cross section and decay width for the coloron and $Z'$ are discussed in detail in Ref.~\cite{Chivukula:2014npa}. Combining them with the mass yields the color discriminant variables for the coloron,
	\bea
		\dcol^C &=&  \frac{16}{3} \left( W_u + W_c\right)
		\Bigg[
				 \frac{g_{C}^{u\,2}}{g_{C}^{u\,2} +  g_{C}^{d\,2} }
				+
				\left( 1 - \frac{g_{C}^{u\,2}}{g_{C}^{u\,2} +  g_{C}^{d\,2} } \right)  \left(\frac{ W_d + W_s }{ W_u + W_c }\right)
			\nonumber
			\\
			&&\qquad	+
				\frac{g_{C}^{b\,2}}{g_{C}^{u\,2} +  g_{C}^{d\,2} }  \left(\frac{W_b }{ W_u + W_c }\right)
		\Bigg]
		\times
		\left\{
		\frac{
			2
			}
			{
				\left( 2 +
				\frac{g_{C}^{t\,2}}{ g_{C}^{u\,2} +  g_{C}^{d\,2} }+ \frac{g_{C}^{b\,2}}{g_{C}^{u\,2} +  g_{C}^{d\,2} }
				\right)^2
			}
		\right\}
	\label{eq:dcolcol-nonu}
	\eea
and for the $\zp$,
\bea
		\dcol^\zp &=&  \frac{2}{3} \left( W_u + W_c\right)
		\Bigg[
				  \frac{g_{\zp}^{u\,2}}{g_{\zp}^{u\,2} + g_{\zp}^{d\,2} }
				+
				\left( 1 -  \frac{g_{\zp}^{u\,2}}{g_{\zp}^{u\,2} + g_{\zp}^{d\,2} } \right)  \left(\frac{ W_d + W_s }{ W_u + W_c }\right)
			\nonumber
			\\
			&&\qquad	+
				\frac{g_{\zp}^{b\,2}}{g_{\zp}^{u\,2} + g_{\zp}^{d\,2} }   \left(\frac{W_b }{ W_u + W_c }\right)
		 \Bigg]
			\times
			\left\{
			\frac{
			2
			}
			{
				\left(
				2
			+ \frac{g_{\zp}^{t\, 2}} {g_{\zp}^{u\,2} + g_{\zp}^{d\,2}}  + \frac{g_{\zp}^{b\, 2}}{g_{\zp}^{u\,2} + g_{\zp}^{d\,2}}
				\right)^2
			}
			\right\}
	\label{eq:dcolzp-nonu}
	\eea
where parts related to resonance production are grouped within the square brackets, while those related to decay are grouped within curly braces. 

The relative strength with which the vector boson couples to the $u$- and $d$-type quarks of the light SM generations,  $g_u^{2} / \left( g_u^{2} + g_d^{2} \right)$, which we call the ``up ratio'', is not accessible by experiments available in the dijet channel. However, equivalent information for quarks in the third generation can be measured by comparing the dijet and heavy flavor cross sections. As detailed in Ref.~\cite{Chivukula:2014npa}, we have
	\beq
		 \frac{g_{t}^{2}} {g_{u}^{2} + g_{d}^{2}} = 2 \frac{\sigma_{t\bar{t}}^V}{\sigma_{jj}^V} \equiv {\rm ``top\ ratio''} \quad\quad
			 \frac{g_{b}^{2}} {g_{u}^{2} + g_{d}^{2}} = 2 \frac{\sigma_{b\bar{b}}^V}{\sigma_{jj}^V} \,, \equiv {\rm ``bottom\ ratio''} \,.
	\eeq
Supplementary measurements of these ratios of cross sections will help pinpoint the structure of couplings of the resonance.

Two items are worth noting at this point:  First, as detailed in Ref.~\cite{Chivukula:2014npa}, unless the resonance has a {\it much} stronger coupling to the $b$ than to quarks in first two generations, the precise strength of the couplings to third-generation quarks becomes relevant to $\dcol$ only through the decay part of the expressions (\ref{eq:dcolcol-nonu}) and  (\ref{eq:dcolzp-nonu}), the part in curly braces.  Second, as we will demonstrate below, the experimentally inaccessible parameter that we call the up ratio appears unlikely to leave us confused as to whether a new dijet resonance is a coloron or a leptophobic $\zp$.

%%%%%%%%%%%%%%%%%%%%%%%%%%%%
%***************************
%***************************
%%%%%%%%%%%%%%%%%%%%%%%%%%%%

\section{Accessible Vector Boson Parameter Space at the 14 TeV LHC}
\label{sec:paramspace}
We now find the region of parameter space in which $\dcol$ can distinguish vector bosons of different color representations. In this region the resonance has not already been excluded by the current searches, is within the reach of future searches, and has a total width that is measurable and consistent with the designation ``narrow''.

The limits on the production cross section times branching ratio ($\sigma \times Br(jj)$) from the (null) ATLAS and CMS~\cite{ATLAS:2012pu, Chatrchyan:2013qha, CMS:kxa} at $\sqrt{s} = 8\,\tev$ searches for narrow resonances establish bounds on dijet resonances. We use the most stringent constraint, from CMS~\cite{CMS:kxa}. As the exclusion limit is provided as $\sigma\times Br(jj) \times (\mathrm{Acceptance})$, we estimate the acceptance for each value of the resonance mass by comparing, within the same theoretical model, $\sigma \times Br(jj)$ as calculated by us vs. $\sigma \times Br(jj) \times (\mathrm{Acceptance})$ provided by CMS. The acceptance is a characteristic of properties of the detector and kinematics, the latter being the same for coloron and $\zp$ to leading order; thus we use throughout our analysis the acceptance deduced from such a comparison  made within a sequential $\zp$ model. The excluded region of parameter space is displayed in gray in Fig. \ref{fig:param_space_col}.

Sensitivity to a dijet resonance in future LHC experiments with $\sqrt{s}=14\,\tev$ depends on the knowledge of QCD backgrounds, the measurements of dijet mass distributions, and statistical and systematic uncertainties. CMS
~\cite{Gumus:2006mxa} has estimated the limits on $\sigma \times Br(jj) \times (\mathrm{Acceptance})$ that will be required in order to attain a $5\sigma$ discovery at CMS with integrated luminosities up to $10\,\ifb$, including both statistical and systematic uncertainties. We obtain the acceptance for CMS at $\sqrt{s} = 14\,\tev$ as described above. The sensitivity for the dijet discovery from $10\,\ifb$ is then scaled to the integrated luminosities $\mathcal{L} = 30,\,100,\,300,\,1000\,\ifb$. The predicted discovery reaches for these luminosities are shown in varying shades of blues for coloron and greens for $\zp$ in Fig. \ref{fig:param_space_col}. 

The total decay width also constrains the absolute values of the coupling constants. On the one hand, experimental searches are designed for narrow-width dijet resonances: $\Gamma/M < 0.15$ ~\cite{Bai:2011ed, Haisch:2011up, Harris:2011bh}. On the other, widths smaller than the experimental dijet mass resolution, $M_\mathrm{res}$, cannot be measured. The region of parameter space that meets both constraints and is relevant to our analysis is shown in Fig. \ref{fig:param_space_col} as the region between the two dashed horizontal curves labeled $\Gamma \ge 0.15 M$ and $\Gamma \le M_\mathrm{res}$. Regions where the width is too broad or too narrow are shown with a cloudy overlay to indicate that they are not accessible via our analysis.

\begin{figure}[t]
{
\includegraphics[width=0.31\textwidth, clip=true]{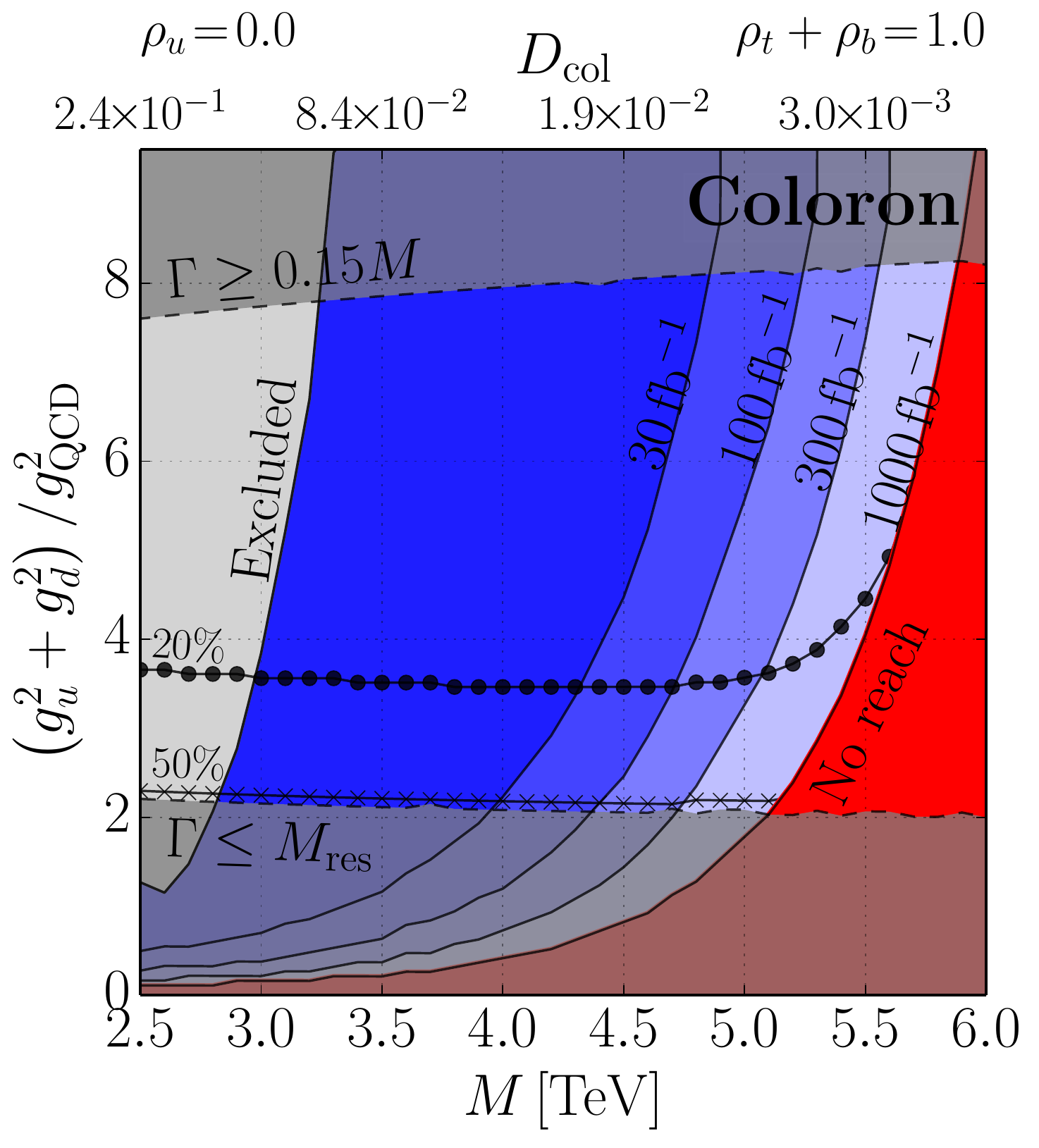}
\includegraphics[width=0.31\textwidth, clip=true]{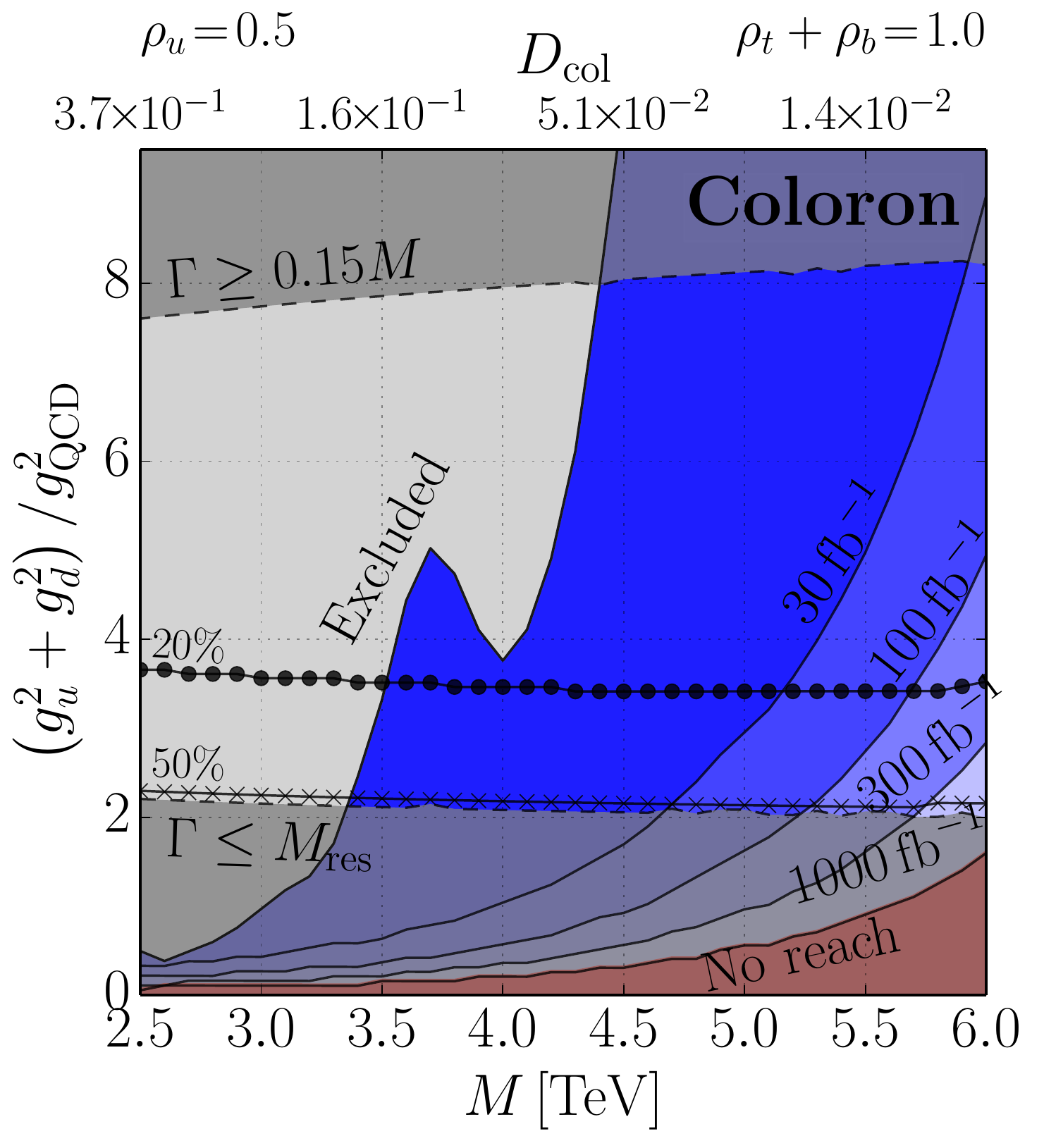}
\includegraphics[width=0.31\textwidth, clip=true]{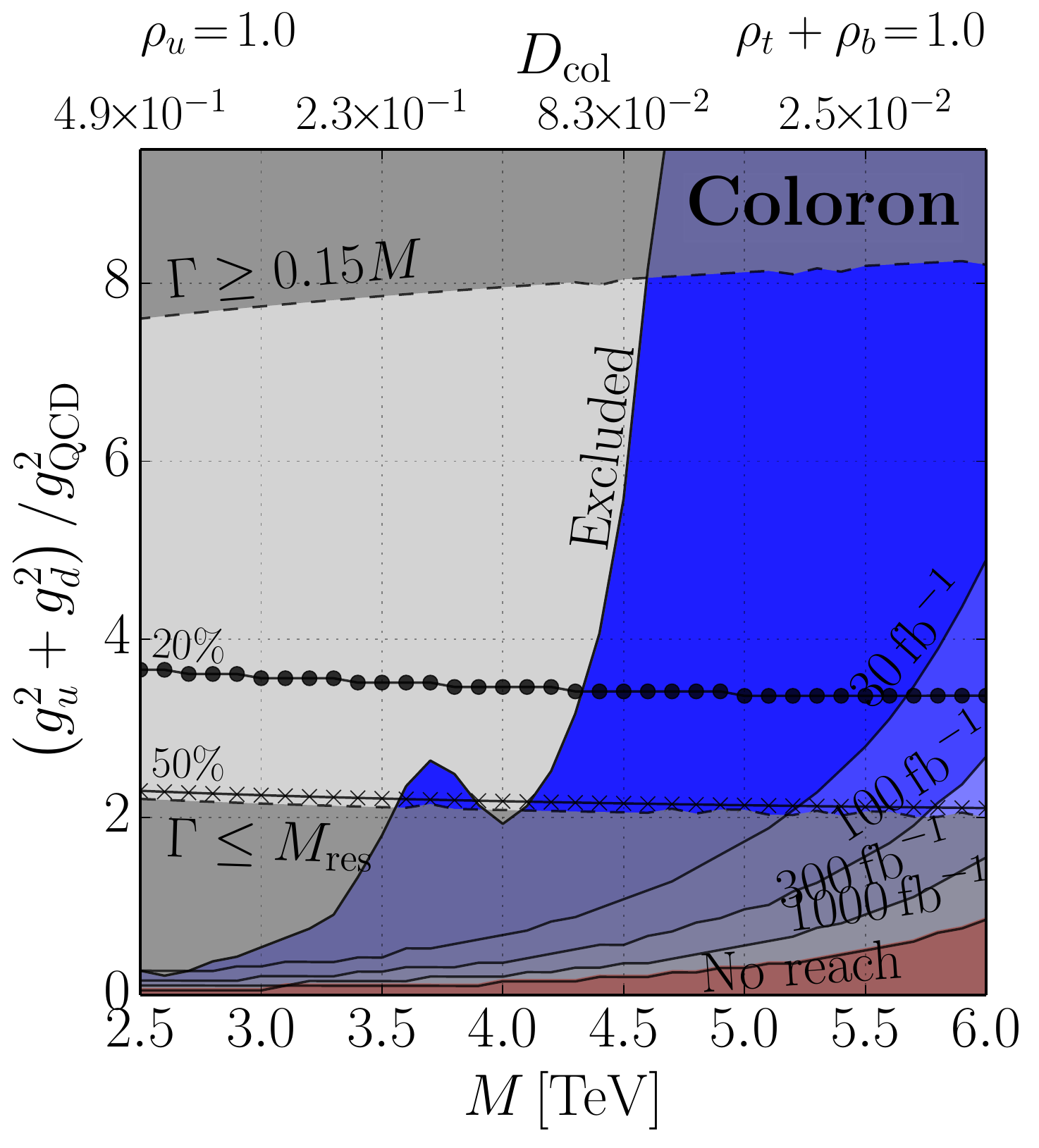}
\\
\includegraphics[width=0.31\textwidth, clip=true]{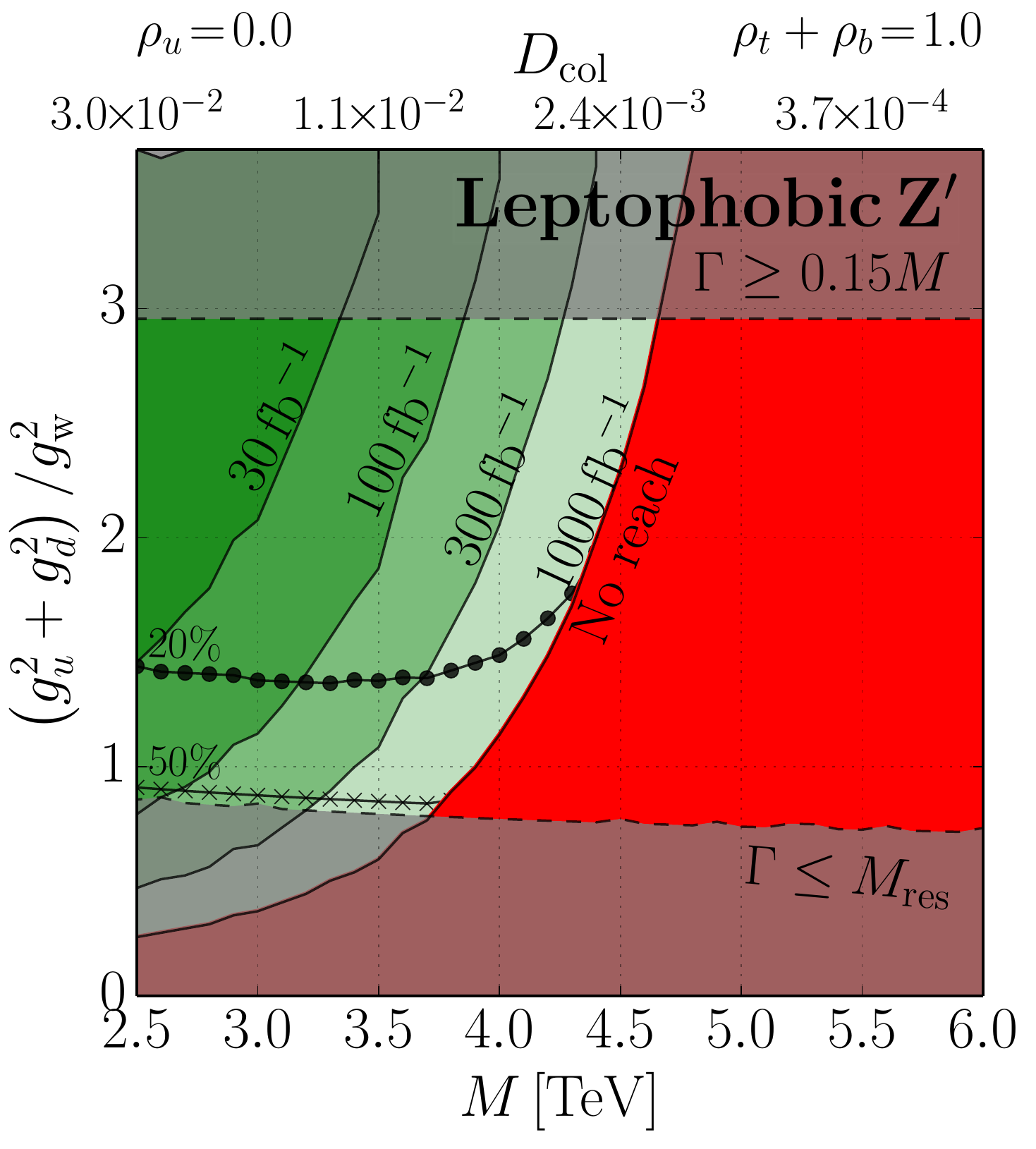}
\includegraphics[width=0.31\textwidth, clip=true]{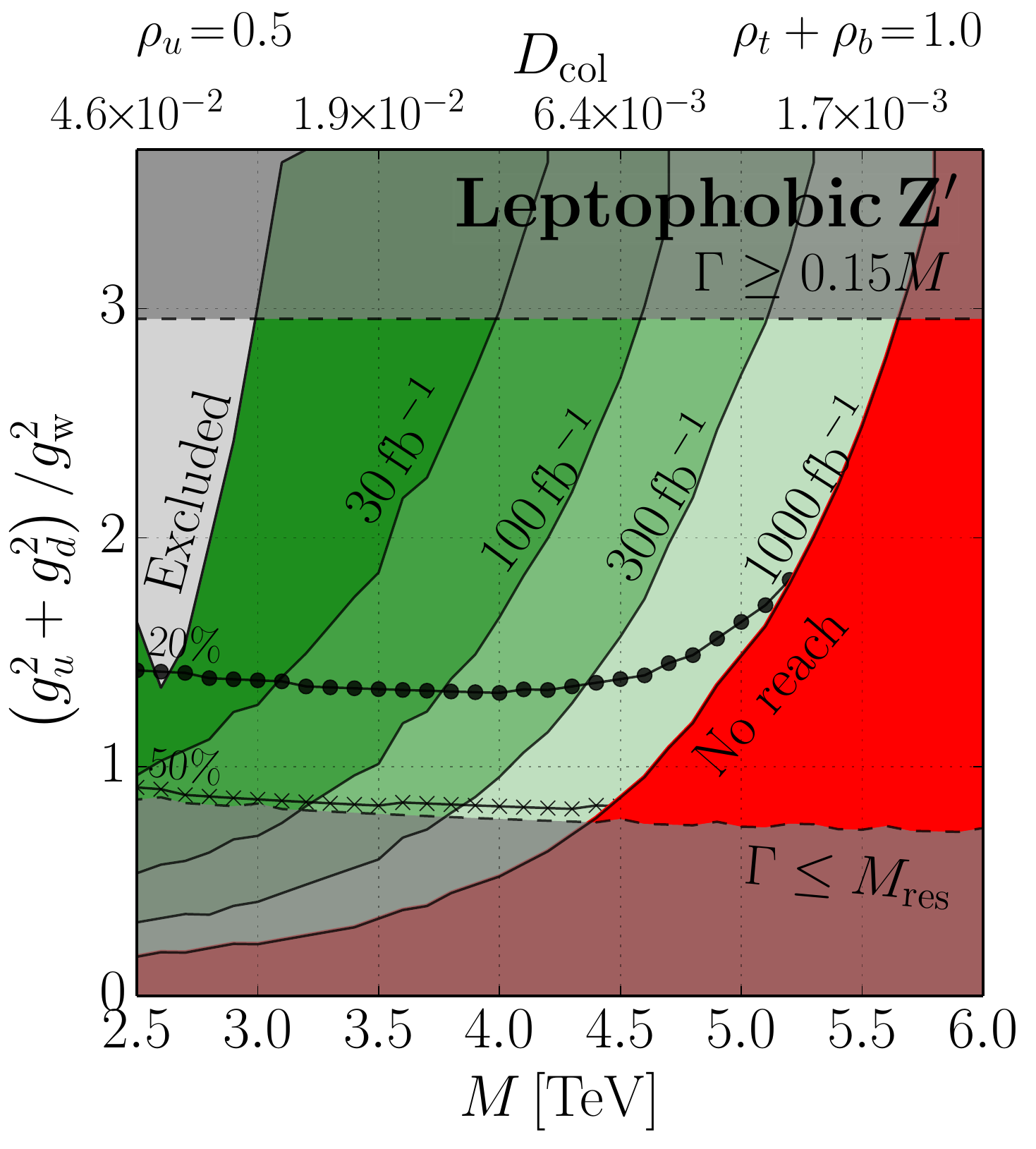}
\includegraphics[width=0.31\textwidth, clip=true]{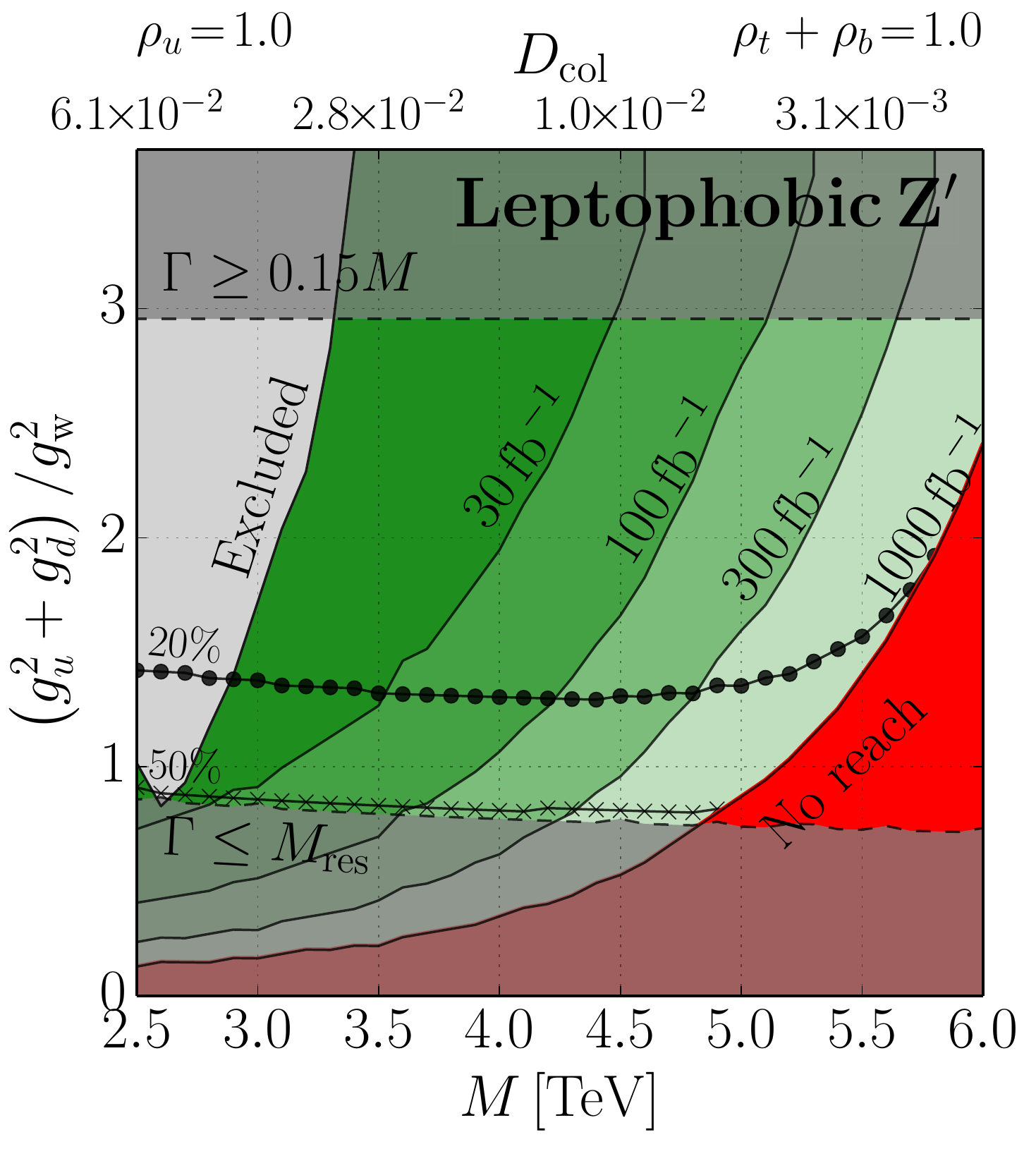}
}
\caption{
Parameter space for colorons (upper row) and $Z'$ bosons (lower row) 
where the color discriminant variable analysis applies at the LHC-14, for different sets of coupling ratios, $\rho_q \equiv {g_q^2}/[{g_u^2+g_d^2}]$; for a wider array of coupling ratios, see Ref.~\cite{Chivukula:2014npa}.
The  5$\sigma$ discovery reach, with statistical and systematic uncertainties included, is shown in varying shades of blue (green) for luminosities ranging from 30 $\ifb$ to 1000 $\ifb$. The red area marked ``no reach'' lies beyond the discovery reach at 1000 $\ifb$ The gray area marked ``excluded'' has been excluded~\cite{CMS:kxa} by LHC-8.  Above the dashed line marked $\Gamma \ge 0.15M$, the narrow-width approximation is not valid. Below the horizontal dashed line marked $\Gamma \le \mres$, the experimental mass resolution is larger than the intrinsic width. In each figure, for fixed $\rho_q$, $\dcol$ is a function of resonance mass only, with values shown along the upper horizontal axis. The contours marked $20\%$ and $50\%$ indicate the region above which the uncertainty in measuring $\dcol$, as estimated in Ref.~\cite{Chivukula:2014npa}, is lower than $20\%$ and $50\%$, respectively. 
}
\label{fig:param_space_col}
\end{figure}

%%%%%%%%%%%%%%%%%%%%%%%%%%%%
%***************************
%***************************
%%%%%%%%%%%%%%%%%%%%%%%%%%%%

\section{Applying $\dcol$ to Flavor-non-universal Vector Bosons}
\label{sec:result-dcols}
We now illustrate how the color discriminant variable $\dcol$ can distinguish whether a  dijet resonance is a coloron or a leptophobic $\zp$ even if it is flavor non-universal.  

\subsection{$C$ and $\zp$ lie in different regions of coupling ratio space}

The value of $\dcol$ at a fixed mass and dijet cross section may correspond to a variety of combinations of values of the three ratios of couplings, the up ratio ($\frac{g_u^2}{g_u^2+g_d^2}$), the top ratio ($\frac{g_t^2}{g_u^2+g_d^2}$), and the bottom ratio ($\frac{g_b^2}{g_u^2+g_d^2}$). The last two are directly determined from the measurements of $\sigma_{t\bar{t}}$ and $\sigma_{b\bar{b}}$ while $\dcol$ is relatively insensitive to the first ratio.  We now show that measuring the mass, width, dijet cross-section, $\sigma_{t\bar{t}}$ and $\sigma_{b\bar{b}}$ can definitively identify the color charge of a newly discovered resonance.  We focus on resonances of mass $3\,\tev$ and $4\,\tev$, as we have seen in Fig.~\ref{fig:param_space_col} that most colorons with lower masses are excluded by the current experiments and most $\zp$ bosons with higher masses are not within reach of $1000\,\ifb$ of LHC data.

In Fig.~\ref{fig:3d_dcol_2_5}, we show the region of parameter space of the three coupling ratios (using the observables $\frac{\sigma_{t\bar{t}}}{\sigma_{jj}}$ and $\frac{\sigma_{b\bar{b}}}{\sigma_{jj}}$ in place of the top and bottom ratios, respectively) in which coloron or $\zp$ models (each displayed as a point) with the same mass lead to a certain range of dijet cross-section and $\dcol$. We choose the range for the dijet cross-section to be within 1 standard deviation of the value that allows a $5\sigma$ discovery at luminosity $1000\,\ifb$. We selected  $\dcol$ to be within $50\%$ of the value $3\times 10^{-3}$ for this illustration, as it permits the required measurements to be made for either a coloron or $\zp$. Points in the accessible area of parameter space are highlighted in blue if the discoverable resonance is a $C$ and in green if it is a $\zp$. These points lie in the blue (green) regions of Fig. \ref{fig:param_space_col} for a coloron ($\zp$).

\begin{figure}[h,t]
{
\includegraphics[width=0.46\textwidth, clip=true]{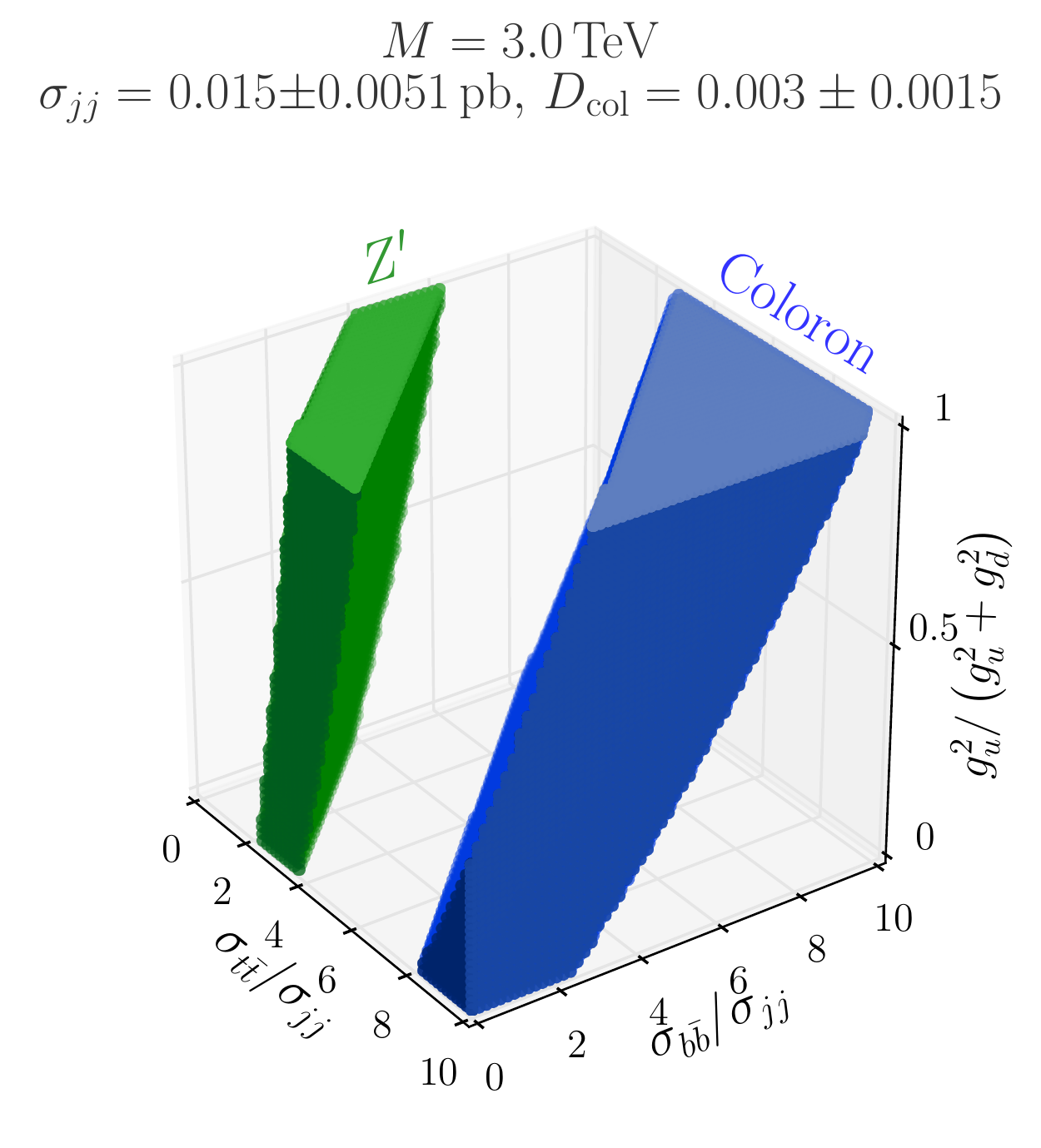}
\includegraphics[width=0.46\textwidth, clip=true]{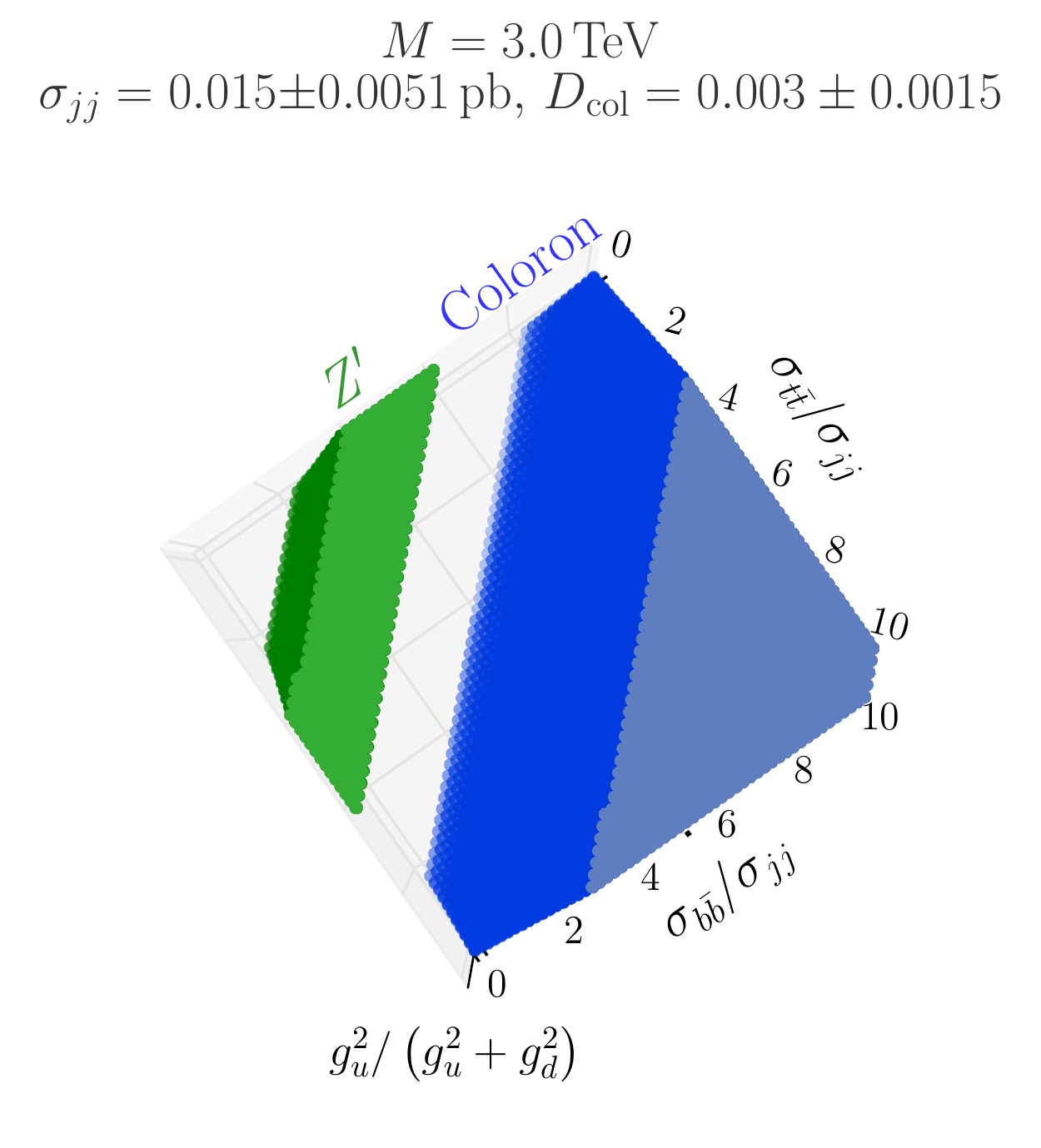}
\\
\includegraphics[width=0.46\textwidth, clip=true]{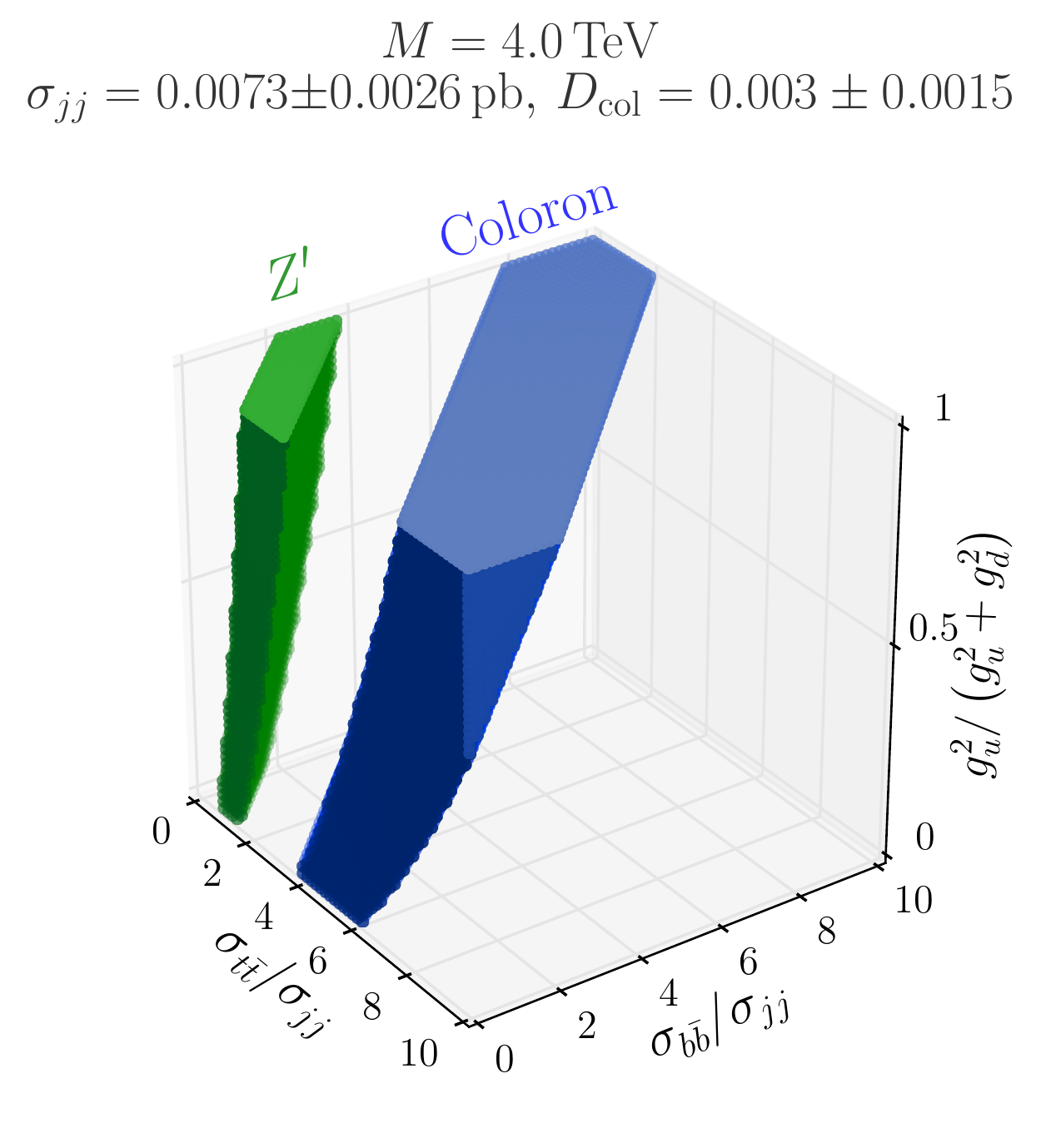}
\includegraphics[width=0.46\textwidth, clip=true]{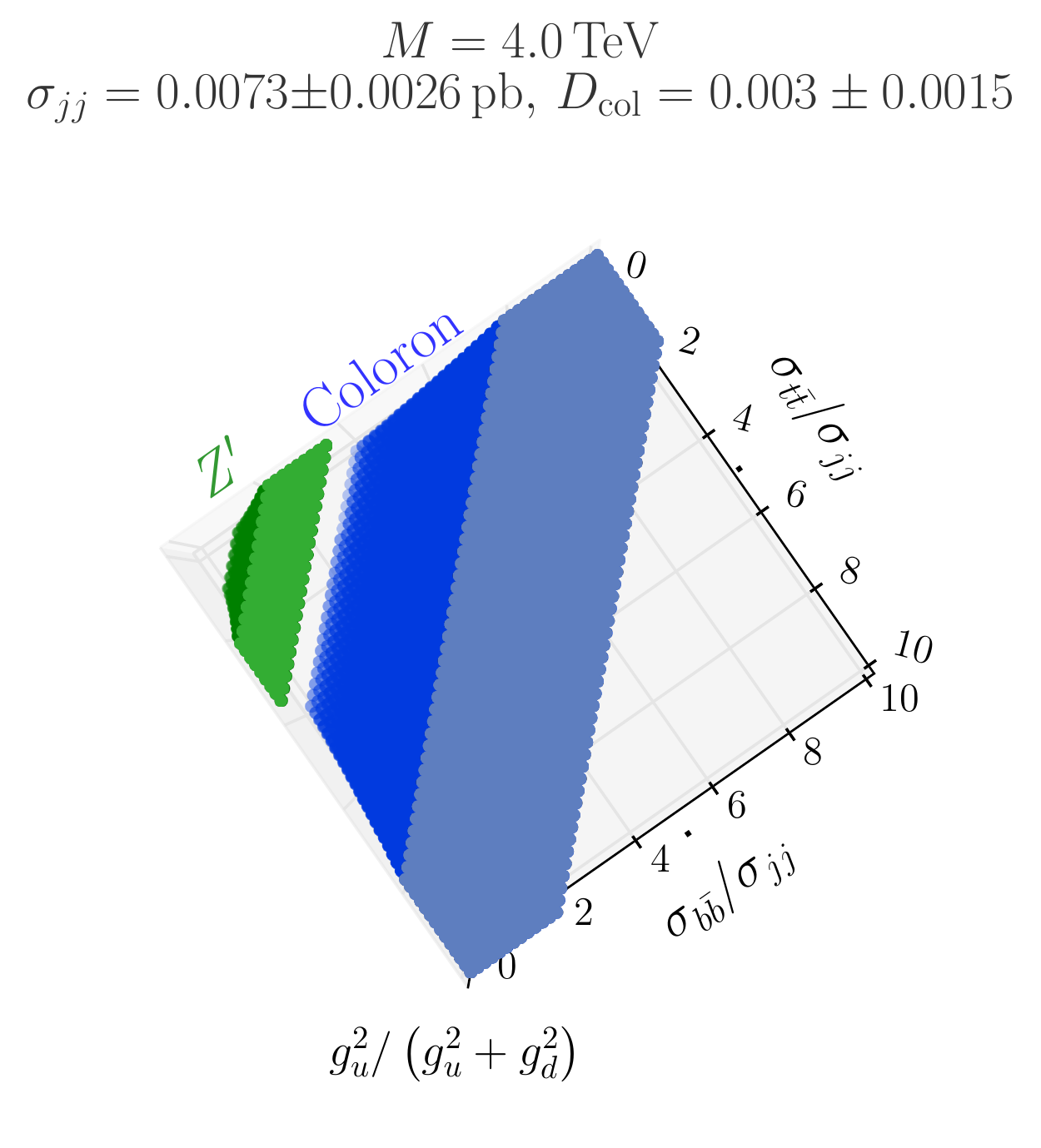}
}
\caption{
Illustration that the proposed measurements suffice to identify the color structure of a new dijet resonance. These plots show regions of the 3-d parameter space $\frac{g_u^2}{g_u^2+g_d^2}$ vs. $\frac{\sigma_{t\bar{t}}}{\sigma_{jj}}$ vs. $\frac{\sigma_{b\bar{b}}}{\sigma_{jj}}$, at fixed values of mass ($3\,\tev$ on the top panels and $4\,\tev$ on the bottom panels) where dijet cross section and $\dcol$ fall within a certain range. The cross section lies within about $35\%$ of the value required for a $5\sigma$ discovery at the LHC with $\sqrt{s} = 14\,\tev$ at $\mathcal{L} = 1000\,\ifb$. Similarly, $D_{col}$ is chosen to lie within $\pm 50\%$ of $3\times 10^{-3}$ for illustration. Points in parameter space that are accessible to the LHC are circles highlighted in blue for $C$ and triangles highlighted in green for $\zp$. Each plot on the left panel is shown again, as viewed from above, on the right. The two views make clear that colorons and Z' bosons lie in distinguishably separate regions of parameter space.
}
\label{fig:3d_dcol_2_5}
\end{figure}

The 3-dimensional plots in the left panel of Fig.~\ref{fig:3d_dcol_2_5} show that colorons and leptophobic $\zp$ which correspond to measurable observables appear in different regions of $\frac{\sigma_{t\bar{t}}}{\sigma_{jj}} \,\mathrm{vs}\, \frac{\sigma_{b\bar{b}}}{\sigma_{jj}} \,\mathrm{vs}\,  \frac{g_u^2}{g_u^2+g_d^2}$ three-dimensional parameter space. The symmetry between the $\frac{\sigma_{t\bar{t}}}{\sigma_{jj}}$ and $\frac{\sigma_{b\bar{b}}}{\sigma_{jj}}$ axes illustrates the rarity of having a heavy resonance produced via $b\bar{b}$ annihilation - the only process in which $b$ quarks could contribute to $\dcol$ without a corresponding contribution from $t$ quarks. In addition, we see that while the up ratio is experimentally inaccessible, the top-view figures displayed in the right panels show that this does not typically lead to confusion between a coloron and a $\zp$. Even having projected the 3D data for all up ratio values onto the bottom-ratio vs. top-ratio plane, the blue coloron and green $\zp$ points still lie in distinct regions.

\subsection{Heavy flavor measurements can separate $C$ from $\zp$}

Fig.~\ref{fig:3d_dcol_2_5} not only shows that the coloron and $\zp$ lie in different regions of parameter space, but also implies that measurements of the top and bottom decays  of the new resonance will almost always enable us to determine its color structure.
Let us assume that a new resonance has been found and that its mass, dijet cross-section, and $\dcol$ have been measured. For illustration, take the values of these observables to be those used in Fig.~\ref{fig:3d_dcol_2_5} (for the same LHC energy,  luminosity and estimated uncertainties). Because there is a gap between the $\zp$ and $C$ regions of that figure when the up ratio is $0$, and because the boundaries of the regions are angled rather than vertical, we can see that a $\zp$ with the minimum up ratio value of $0$ would not be mistaken for a coloron.  But a $\zp$ with the maximum up ratio (equal to $1$) lies as close as possible to the coloron region of parameter space. So, to see how close the two regions can get, we should compare a $\zp$ with an up ratio of $1$ (one that does not couple to $d$ or $s$ quarks) to a coloron with varying values of the up ratio.

This very comparison is presented in Fig.~\ref{fig:2d_dcol}, which is plotted in the top-ratio vs bottom-ratio plane. The up ratio for the $\zp$ is fixed to be $1$; the up ratio for the $C$ is varied from $0$ (left panels) to $1$ (right panels). As the up ratio for the coloron increases, the blue coloron band of parameter space moves out from the origin, away from the green $\zp$ band of parameter space. Conversely, if we decreased the up ratio of the $\zp$ boson, the green $\zp$ band would shift closer to the origin, away from the blue coloron band. In general, the coloron and $\zp$ regions do not overlap.

Given the shape and orientation of the regions corresponding to color-singlet and color-octet resonances in the plots, measuring both the top ratio and bottom ratio would clearly allow us to distinguish the new resonance's color structure. Moreover, we see that if either the top ratio or bottom ratio were measured to be sufficiently large, we would know that the resonance must be a coloron (because the $\zp$ region is already at its maximum distance from the origin). For example, a measurement of $\sigma_{t\bar{t}}/\sigma_{jj} \gtrapprox 6$ for a $3\,\tev$ resonance or $\sigma_{t\bar{t}}/\sigma_{jj} \gtrapprox 3$ for a $4\,\tev$ resonance, for the values of $\sigma_{jj}$ used in Fig.~\ref{fig:2d_dcol}, would identify it as a color-octet.

There could still be a rare situation where our inability to measure the up ratio would prevent us from determining the color structure of a new resonance. The regions of parameter space corresponding to the extreme cases of a coloron with only down-type light quark couplings and a $\zp$ with only up-type light quark couplings could potentially overlap. As discussed in Ref.~\cite{Chivukula:2014npa}, this is more likely to happen for heavier resonances.. For example, given our estimates of uncertainties, such an overlap could potentially occur for a $4\,\tev$ resonance as illustrated by the close approach of the $\zp$ and coloron bands in the lower left panel of Fig.~\ref{fig:2d_dcol}.

Determining the color structure of a resonance generally requires measurements of both $\sigma_{t\bar{t}}$ and $\sigma_{b\bar{b}}$. As with the dijet cross sections, systematic uncertainties for these measurements will be obtained after the experiment (at $14\,\tev$) has started.  Our results illustrate~\cite{Chivukula:2014npa} that measuring the $t\bar{t}$ and $b\bar{b}$ cross sections to an uncertainty of $\mathcal{O}(1)$ would still provide significant information.

\begin{figure}[t]
{
\includegraphics[width=0.99\textwidth, clip=true]{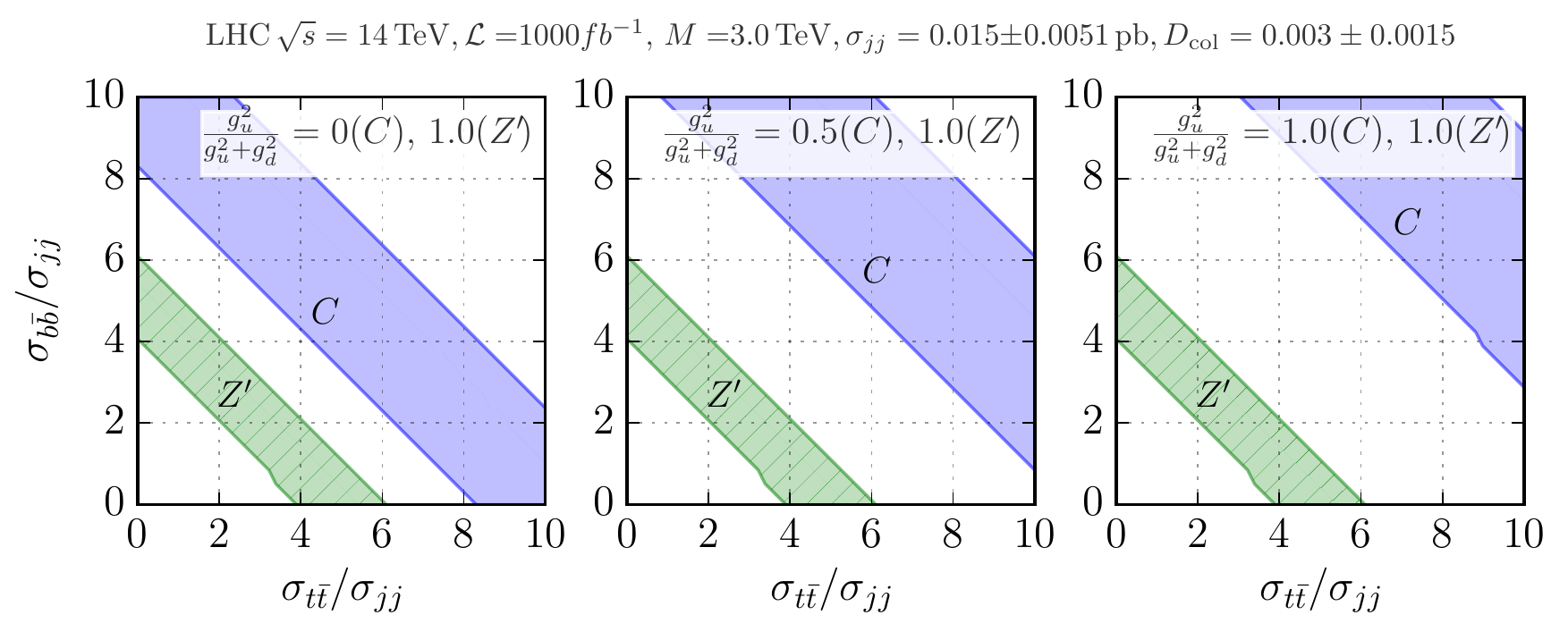}
\\
\includegraphics[width=0.99\textwidth, clip=true]{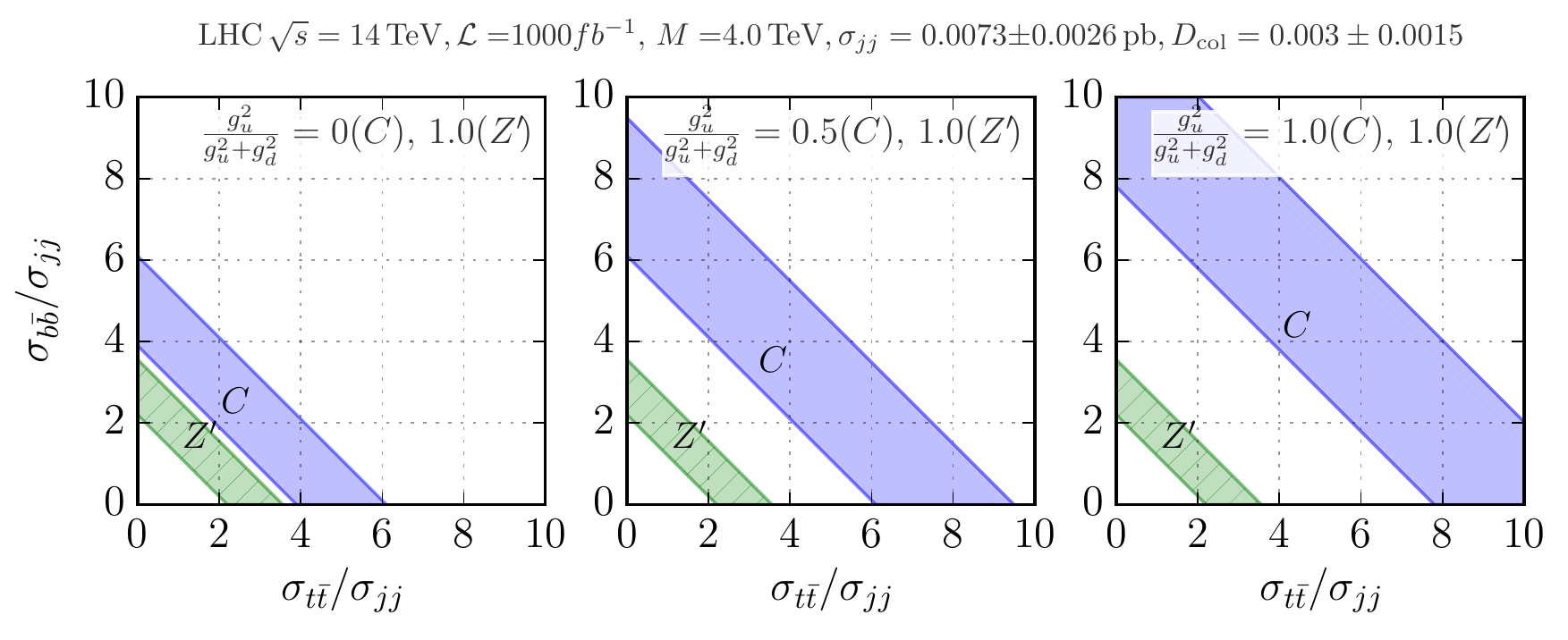}
}
\caption{Measuring $\sigma_{t\bar{t}}$ and $\sigma_{b\bar{b}}$ will show whether a resonance of a given mass, dijet cross-section, and $\dcol$ is a coloron or $\zp$. We display regions of parameter space corresponding to a dijet cross section within $1\sigma$ uncertainty of the value required for a discovery at the LHC with $\sqrt{s} = 14\,\tev$ at $\mathcal{L} = 1000\,\ifb$, and that have the color discriminant variable within a range of $50\%$ around the value of $\dcol = 3\times 10^{-3}$, for resonances with masses $3\,\tev$ (top panel) and $4\,\tev$ (bottom panel). Coloron and $\zp$ models within reach are, respectively, displayed in a blue region and a diagonally-hatched green region.  The colorons and $Z'$ bosons lie in different regions of parameter space.  However, at the bottom left panel for a 4 TeV resonance, the bands for a $C$ coupling only to down-type light quarks and a $\zp$ coupling only to up-type light quarks approach closely.
}
\label{fig:2d_dcol}
\end{figure}

%%%%%%%%%%%%%%%%%%%%%%%%%%%%
%***************************
%***************************
%%%%%%%%%%%%%%%%%%%%%%%%%%%%

\section{Using $\dcol$ to Separate Vectors, Scalars, and Fermions}
\label{sec:vector-scalar-fermion}

We now report on work~\cite{Chivukula:2014pma} using the color discriminant variable to compare resonances that decay to $\bar{q}q$, $qg$, and $gg$ final states. We consider three compelling benchmark scenarios to describe the different di-jet resonances: the flavor universal coloron model \cite{Simmons:1996fz, Chivukula:1996yr} for $\bar{q}q$ resonances, the excited quark model of Ref. \cite{Baur:1989kv, Baur:1987ga} for $qg$ resonances and the general parameterization in \cite{Han:2010rf} of color-octet scalar interactions for $gg$ resonances. 
All of the results are shown in the relevant mass-coupling parameter space that is both not excluded by the 8 TeV LHC analyses \cite{Aad:2014aqa, CMS:kxa, Chatrchyan:2013qha} and conducive to a 5$\sigma$ discovery of the resonance in the di-jet channel at the 14 TeV LHC. The LHC-8 excluded regions are extracted from the ATLAS \cite{Aad:2014aqa} and CMS \cite{CMS:kxa, Chatrchyan:2013qha} searches; the LHC-14 discovery reach is evaluated based on Monte Carlo simulations. 

\subsection{Benchmark Models}

Let us review the essential features of our three benchmark models.  The flavor universal coloron model was described above.  The other two will be sketched here.

\subsubsection{Excited quarks ($\mathbf{q^*}$)}

Quark-gluon resonances are a general prediction of composite models with excited quarks \cite{Baur:1989kv, Baur:1987ga}.  We will focus on the phenomenological model of \cite{Baur:1989kv}, which describes an electroweak doublet of excited color-triplet vector-like quarks $q^{*}=(u^{*}, d^{*})$ coupled to first-generation ordinary quarks. In this model, 
right-handed excited  quarks interact with gauge bosons and ordinary (left-handed) quarks through magnetic moment interactions described by  the effective Lagrangian:
 \begin{align}
 \begin{split}\label{eq:L_qStar}
& \mathcal{L}_{int}=\frac{1}{2\Lambda} \bar{q}^{*}_R \sigma^{\mu\nu} \left[ g_S f_S \frac{\lambda^{a}}{2} G^{a}_{\mu\nu}+g f \frac{\mathbf{\tau}}{2}\cdot \mathbf{W}_{\mu\nu} + g^{'} f^{'} \frac{Y}{2} B_{\mu\nu} \right]  q_L + \text{H.c}. 
 \end{split}
 \end{align}

The excited quarks can decay into $qg$ or into a quark plus a gauge boson. The corresponding decay rates are:
 \begin{align}
 \begin{split}\label{eq:Gamma_qStar}
 &\Gamma(q^{*}\to qg)=\frac{1}{3}\alpha_S f^2_{S}\frac{m^3_{q*}}{\Lambda^2} \qquad\qquad
 \Gamma(q^{*}\to q\gamma)=\frac{1}{4}\alpha f^2_{\gamma}\frac{m^3_{q*}}{\Lambda^2} \\ 
 &\Gamma(q^{*}\to qV))=\frac{1}{8}\frac{g^2_V}{4\pi}f^2_{V}\frac{m^3_{q*}}{\Lambda^2}\left[ 1-\frac{m^2_V}{m^2_{q*}}\right]^2\left[ 2+\frac{m^2_V}{m^2_{q*}}\right]
 \end{split}
 \end{align}
 with $V=W,Z$ and with the definitions

\begin{equation}
\label{eq:f_qStar}
f_{\gamma}=f T_3 +f'\frac{Y}{2}\qquad  f_{Z}=f T_3\cos^2\theta_W -f'\frac{Y}{2}\sin^2\theta_W \qquad
   f_{W}=\frac{f}{\sqrt{2}}~.
 \end{equation}
The $q^{*}\to qg$ branching ratio is about 0.8 for $f_S=f=f'$.

Excited quarks are singly produced at the LHC through quark-gluon annihilation and, as just noted, they dominantly decay into $qg$. For our analysis, we choose the benchmark parameters $\Lambda=m_{q^{*}}$ and $f_S=f=f'$, while allowing the overall coupling strength to vary. By way of comparison, recent LHC searches, CMS \cite{CMS:kxa, Chatrchyan:2013qha} and ATLAS \cite{Aad:2014aqa} have used the same value of $\Lambda$ with $f_S=f=f'=1$.

\subsubsection{Color-octet scalars ($\mathbf{S_8}$)}

A gluon-gluon final state can generally arise from decay of colored scalars in models with extended color gauge structures \cite{Hill:1991at, Frampton:1987dn, Martynov:2009en, Chivukula:2013hga, Bai:2010dj}. 
In this work we adopt the general effective interaction for a color octet scalar, $S_8$, introduced in \cite{Han:2010rf}:
\begin{equation}\label{eq:L-s8}
\mathcal{L}_{S_8}=g_S d^{ABC}\frac{k_S}{\Lambda_S} S^A_8 G^B_{\mu\nu}G^{C, \mu\nu} \ ,
\end{equation}
where $d$ is the QCD totally symmetric tensor.

A colored scalar of this kind is singly-produced at the LHC through gluon-gluon annihilation. We consider the case in which it decays entirely (or almost entirely) into gluons. The corresponding decay rate reads:
\begin{equation}\label{eq:Gamma-s8}
\Gamma(S_8)=\frac{5}{3}\alpha_S \frac{k^2_S}{\Lambda^2_S}m^3_{S_8}~.
\end{equation}
We set $\Lambda_S=m_{S_8}$ and we present results for different couplings $k_S$.  Similarly, CMS \cite{CMS:kxa, Chatrchyan:2013qha} and ATLAS \cite{Aad:2014aqa} present searches for $\Lambda_S=m_{S_8}$ and $k_S=1$.

\subsection{LHC Discovery Reach}\label{sec:reach}

For each type of dijet resonance, we have derived~\cite{Chivukula:2014pma} the relevant mass and coupling parameter space for our analysis, namely the region that is not yet excluded by LHC-8 analyses and in which a 5$\sigma$ discovery will be possible at the 14 TeV LHC.  Fig.~\ref{fig:excl-disc} summarizes our estimates of the 5$\sigma$ reach at the 14 TeV LHC in the mass-vs.-coupling plane for colorons, excited quarks, and scalar-octets, for integrated luminosities of 30 -- 3000 fb$^{-1}$. The discovery reach we find for the coloron is very similar to those already derived in \cite{Yu:2013wta, Dobrescu:2013cmh} and in \cite{Atre:2013mja}. 

Within each pane of Fig.~\ref{fig:excl-disc}, we may identify a ``region of interest" where a resonance of a given mass and coupling is not excluded by LHC (i.e., is not in the blue region at left), is relatively narrow (lies below the horizontal dashed curve) and would be detectable at LHC-14 at the indicated luminosity (is within the central light-grey region).  The portion of this region of interest that lies above the horizontal dotted curve is accessible to coloron discriminant variable analysis, while the area below the dotted curve region is also accessible to jet energy profile analysis~\cite{Chivukula:2014pma}.
\begin{figure}[h,t]
{
\includegraphics[width=0.46\textwidth, clip=true]{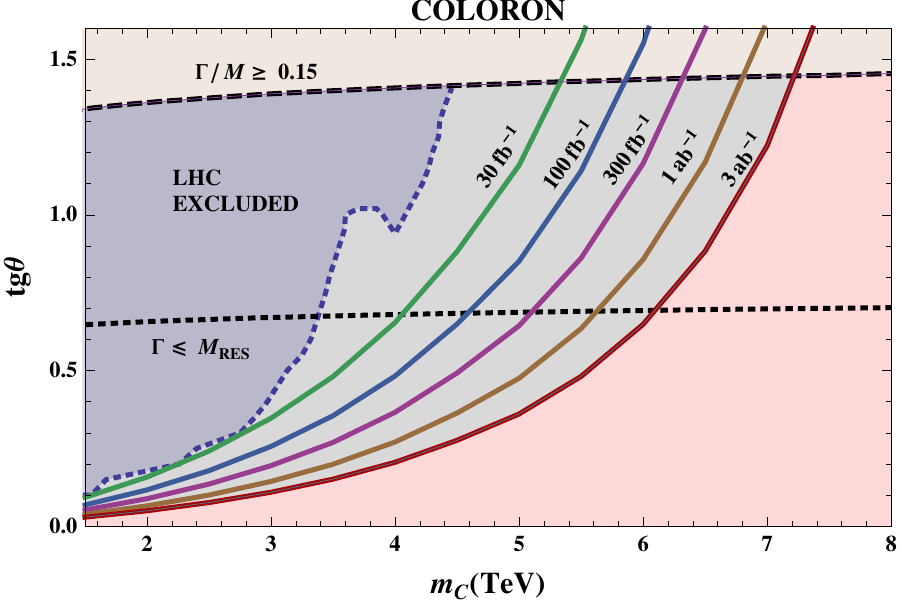}
\includegraphics[width=0.46\textwidth, clip=true]{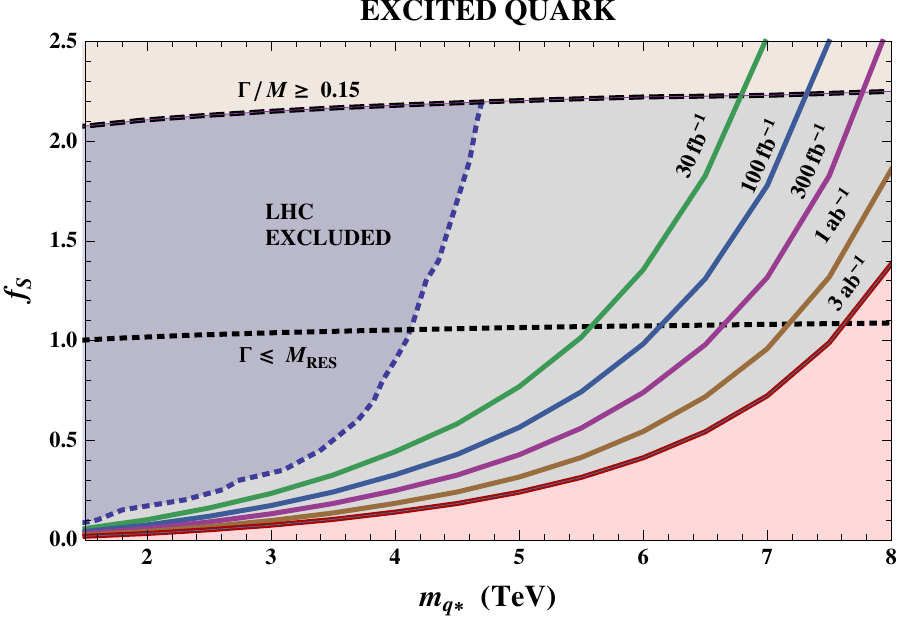}
\includegraphics[width=0.46\textwidth, clip=true]{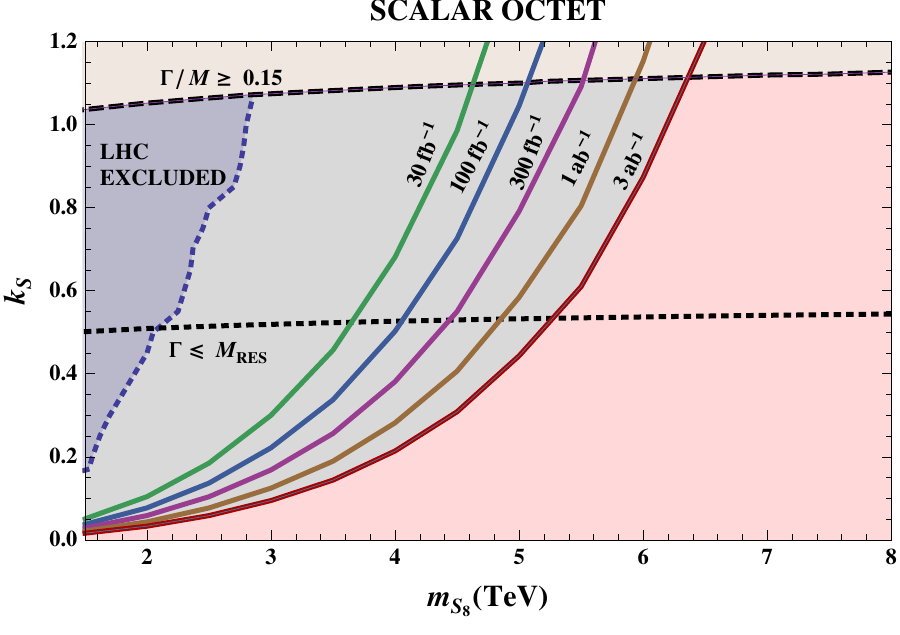}
}
\caption{\small In each pane from left to right: regions of parameter space excluded by LHC-8 (blue), accessible to LHC-14 (pale grey), or inaccessible at LHC-14 (pink). Thick colored curves show the 5$\sigma$ reach at luminosities from 30 -- 3000 fb$^{-1}$. Above the upper dashed line, the resonance is too broad to detect ($\Gamma=0.15 M$); below the lower dotted line it is narrower than the experimental resolution ($\Gamma\leq M_{res}$), where $M_{res} = 0.035 M$ \cite{CMS:kxa}. }
\label{fig:excl-disc}
\end{figure}

\subsection{The Color Discriminant Variable}\label{sec:D-col}
We have evaluated the value of $D_{col}$ for our benchmark di-jet resonances -- flavor universal colorons, excited quarks and scalar octets, in the allowed and accessible range of resonance masses. 
The dependence of $D_{col}$ on the di-jet mass is controlled by the parton distribution functions (PDFs), since the quark and gluon parton content vary with the energy scale of the di-jet process. We calculate the di-jet resonance production cross section by using the CT10 \cite{Lai:2010vv} next-to-leading-order PDF set with factorization and renormalization scales fixed at the resonance mass value.  The measurement of $D_{col}$ at the LHC is affected by the statistical and systematic uncertainties on measurements of the di-jet cross section, the resonance mass and the resonance width. Furthermore, $D_{col}$ is only experimentally accessible if the mass resolution of the detector is less than the intrinsic width of the resonance.  Our analysis\cite{Chivukula:2014pma} has fully included these uncertainties.

Fig. \ref{fig:D} shows the $\log_{10} D_{col}$ values, including the statistical and systematic uncertainties, for the three types of di-jet resonances $q^{*}$, $C$, $S_8$, as a function of the di-jet resonance mass at the 14 TeV LHC for different integrated luminosities. We observe that an excited quark resonance can be efficiently distinguished from either a coloron or a scalar octet resonance by the color discriminant variable at the 14 TeV LHC. Discriminating between colorons and scalar octets using the color discriminant variable is more challenging, but we find it should be possible to establish a separation which ranges from $\sim2\sigma$ at $M\simeq 4$ TeV to $\sim 3\sigma$ at $M\simeq 6$ TeV. 

\begin{figure}[h,t]
{
\includegraphics[width=0.46\textwidth, clip=true]{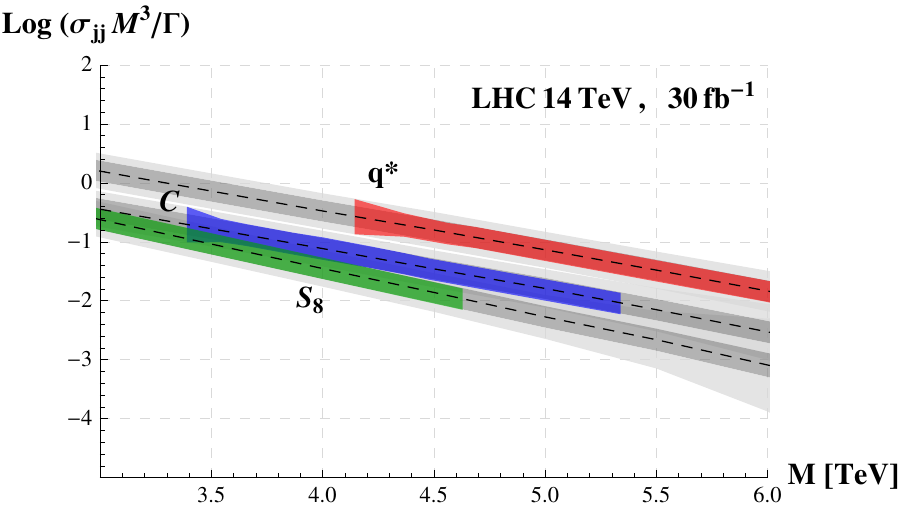}
\includegraphics[width=0.46\textwidth, clip=true]{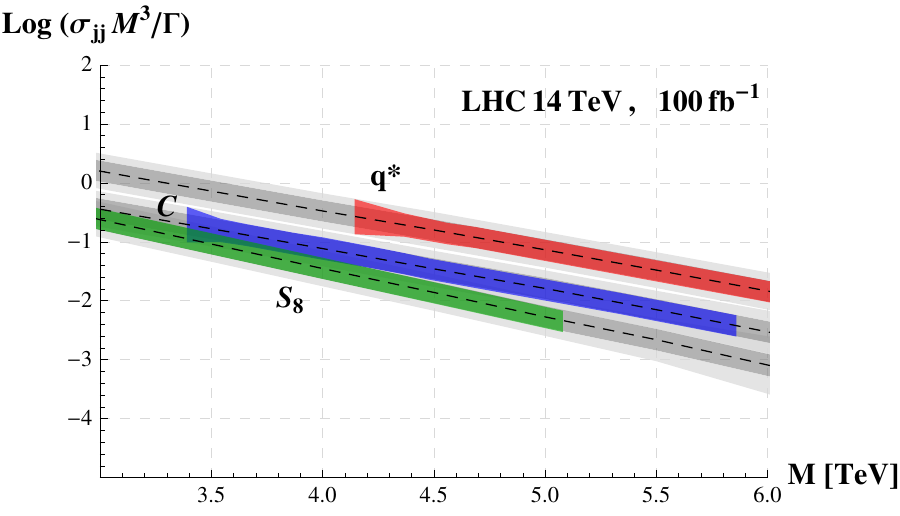}\\
\includegraphics[width=0.46\textwidth, clip=true]{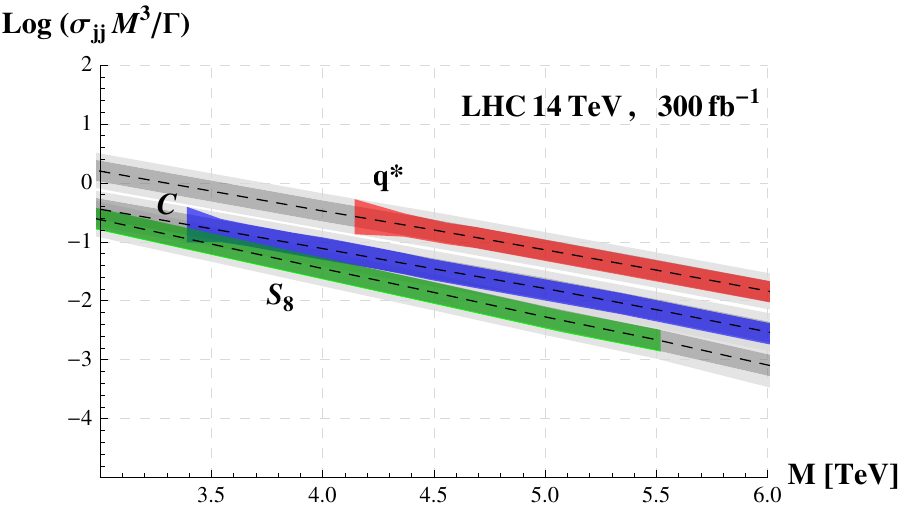}
\includegraphics[width=0.46\textwidth, clip=true]{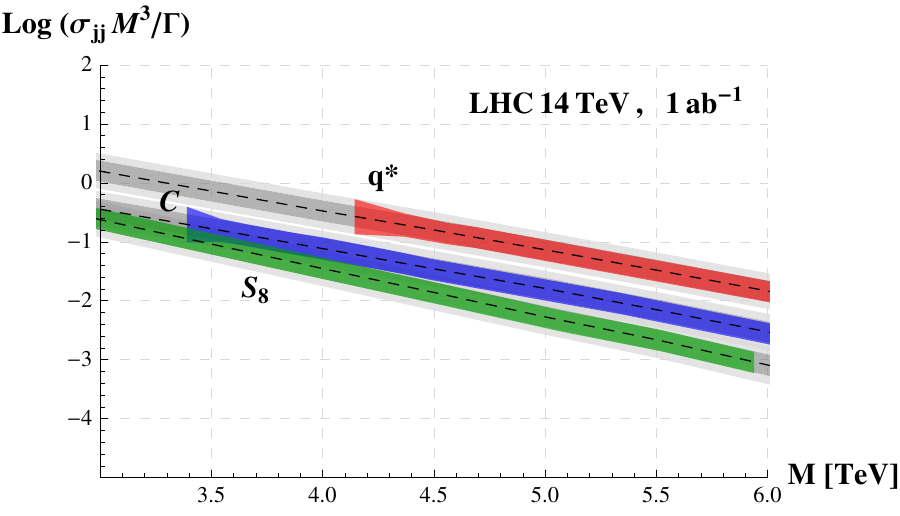}
}
\caption{ \small Each pane shows $\log_{10} D_{col}$ for flavor universal colorons ($C$), excited quarks ($q^*$) and scalar octets ($S_8$) as a function of the dijet resonance mass at the 14 TeV LHC with a particular integrated luminosity. The outer (inner) grey (light grey) bands represent the $\pm 1$-sigma statistical plus systematic uncertainty on $\log_{10} D_{col}$ when the resonance width is narrow (broad): $\Gamma=M_{res}$ ($\Gamma=0.15 M$). The colored bands [red for $q*$, blue for ($C$), green for ($S_8$)] show the $\log_{10} D_{col} \pm 1 \sigma$ values obtained in the region of parameter space where the resonance is allowed by LHC-8 analyses, is neither too broad nor too narrow, and is amenable to discovery at LHC-14. }
\label{fig:D}
\end{figure}

%%%%%%%%%%%%%%%%%%%%%%%%%%%%
%***************************
%***************************
%%%%%%%%%%%%%%%%%%%%%%%%%%%%
\section{Discussion}
\label{sec:discussions}

If the LHC discovers a new dijet resonance, the color discriminant variable can help identify what has been found.  The variable $\dcol$ is constructed from measurements available directly after discovery: namely, the resonance's mass, its total decay width, and its dijet cross section.  This talk has reviewed recent results that apply the color discriminant variable to two new situations: (a) distinguishing between color-singlet and color-octet vector bosons whose couplings to quarks are not flavor-universal and (b) telling apart vector bosons, colored scalars, and excited quarks.

The first analysis assumes the new resonance couples identically to quarks of the first two generations, in which case $\dcol$  depends on three model-specific ratios of coupling constants:  the up ratio (${g_u^2}/[{g_u^2+g_d^2}]$), the top ratio (${g_t^2}/[{g_u^2+g_d^2}]$), and the bottom ratio (${g_b^2}/[{g_u^2+g_d^2}]$). We find the method is generally not dependent on knowing the up ratio, a quantity which is not presently accessible to experiment. Since $\dcol$ is insensitive to chiral structure, discriminating between color-singlet and color-octet resonances with flavor non-universal couplings requires only measurements of the $t\bar{t}$ and $b\bar{b}$ resonance cross sections. 

The second analysis shows that an excited quark can be cleanly distinguished from either a coloron or a color-octet scalar by the color discriminant variable at the 14 TeV LHC. Establishing the distinction between colorons and color-octet scalars using the color discriminant variable is more challenging, but we still find the possibility of a $\sim2(3)\sigma$ separation for resonance masses of order $4(6)$ TeV.

To summarize: we have generalized the color discriminant variable for use in determining the color structure of new bosons that may have flavor non-universal couplings to quarks or for comparing $q\bar{q}$, $qg$ and $gg$ dijet resonances. . We focused on resonances having masses $2.5-6.0\,\tev$ for the LHC with center-of-mass energy $\sqrt{s} = 14\,\tev$ and integrated luminosities up to $1000,\,\ifb$. After taking into account the relevant uncertainties and exclusion limits from current experiment and sensitivity for future experiments, we find that the future runs of the LHC can reliably determine the color structure of a resonance decaying to the dijets.

\section*{Acknowledgments} 
This material is based upon work supported by the National Science Foundation under Grant No. PHY-0854889. We wish to acknowledge the support of the Michigan State University High Performance Computing Center and the Institute for Cyber Enabled Research. PI is supported by Development and Promotion of Science and Technology Talents Project (DPST), Thailand.  EHS and RSC thank the Kobayashi Maskawa Institute for the Origins of Particles and the Universe for hospitality during the SGT15, and also thank the Aspen Center for Physics and the NSF Grant \#1066293 for hospitality during the writing of this Proceedings.

\bibliographystyle{ws-procs975x65}
\bibliography{colvszp}

\end{document}

\end{document}